\documentclass[prl,twocolumn,showpacs,amsmath,superscriptaddress]{revtex4}

\usepackage{graphicx}
\usepackage{dcolumn}
\usepackage{bm}

\begin{document}
\preprint{APS/123-QED}

\title{Fluorescence and phosphorescence from individual  C$_{60}$ molecules\\ excited by local electron tunneling}

\author{Elizabeta \'Cavar}
\author{Marie-Christine Bl\"um}
\author{Marina Pivetta}
\author{Fran\c cois Patthey}
\affiliation{Ecole  Polytechnique F\'ed\'erale  de Lausanne (EPFL), Institut   de Physique des Nanostructures, CH-1015 Lausanne, Switzerland}
\author{Majed Chergui}
\affiliation{Ecole  Polytechnique F\'ed\'erale  de Lausanne (EPFL), Laboratoire de Spectroscopie Ultrarapide, ISIC, CH-1015 Lausanne, Switzerland}
\author{Wolf-Dieter Schneider}
\affiliation{Ecole  Polytechnique F\'ed\'erale  de Lausanne (EPFL), Institut   de Physique des Nanostructures, CH-1015 Lausanne, Switzerland}

\date{\today}%

\begin{abstract}
Using the highly localized current of electrons tunneling through a double barrier Scanning Tunneling Microscope (STM) junction, we excite luminescence from a selected C$_{60}$ molecule in the surface layer of  fullerene nanocrystals  grown on  an ultrathin NaCl film on Au(111).
In the observed luminescence fluorescence and phosphorescence spectra, pure electronic as well as  vibronically  induced transitions of an individual C$_{60}$ molecule are identified, leading to unambiguous chemical recognition  on the single-molecular scale.
\end{abstract}

\maketitle
Light emission induced by electrons tunneling through the junction formed by the sample and the tip of a Scanning Tunneling Microscope (STM) has been proposed to characterize the optical properties of nanoscale objects at surfaces \cite{Gimzewski88}.  Contrary to  conventional non-local techniques, the local character of this method offers the unique possibility  to select and probe individual atoms, molecules or clusters on surfaces.\\
\indent Photon emission due to the decay of localized surface plasmons, excited by inelastic electron tunneling (IET) has been observed on metal surfaces \cite{Berndt91,Berndt93B}, as well as on supported metallic  nanoparticles  \cite{Nilius00}. Luminescence spectra have  been acquired from semiconductor heterostructures  \cite{Alvarado91}, quantum well states of metallic films \cite{Hoffmann01}.
Recently, luminescence from supported molecules has been obtained \cite{Qiu03,Dong04}  by successfully decoupling them from the metallic substrate in order to avoid quenching of the radiative transitions \cite{Berndt93,Dong03},
using either a thin oxide film \cite{Qiu03} or  several molecular layers \cite{Dong04}.\\
\indent However, unambiguous chemical identification of single complex molecules requires the observation and identification of \emph{several} vibrational and/or electronic-vibrational transitions, which are the spectroscopic fingerprint of the species.
Here we present the first observation of energy resolved luminescence from an individually selected C$_{60}$ molecule
excited by electrons tunneling through a double barrier STM junction.
A comparison with the luminescence spectra obtained by non-local laser spectroscopy
 from dispersed C$_{60}$ molecules in rare gas and glass matrices \cite{Sassara96B,Sassara97,Sassara96,Hung96,vandenHeuvel94,vandenHeuvel95}, and from  solid C$_{60}$  \cite{Guss94,vandenHeuvel95B,Akimoto02}  enables us to
  demonstrate the molecular origin of the detected light and to identify the observed spectral features with  pure electronic transitions  and with vibronic  transitions induced via Jahn-Teller (JT) and Herzberg-Teller (HT) coupling \cite{Negri92,Orlandi02}.
The present novel observation of both,  fluorescence (singlet-to-singlet transitions)  \emph{and} phosphorescence (triplet-to-singlet transitions) constitutes a solid basis for the chemical identification of an  individual   C$_{60}$  molecule.\\
\indent C$_{60}$ nanocrystals were grown on NaCl layers deposited onto a Au(111) substrate. NaCl was evaporated from a Knudsen cell on a clean Au(111) surface at room temperature. Subsequently, the C$_{60}$ molecules were sublimated on the NaCl covered substrate.
The experiments were performed with a homebuilt ultrahigh vacuum (UHV) STM operating at a temperature of $50\,$K, using cut PtIr tips.
The photons emitted from the tunnel junction were collected by a lens placed inside the cryostat, guided through an optical system outside the UHV chamber  to the spectrograph, and detected by a CCD camera. The wavelength resolution of the experiment was 8 nm, corresponding to $\approx\,$20 meV in the energy range of interest.
The spectra were acquired with closed feedback loop  while tunneling over a defined position on the sample, e.g. over a single molecule, with a typical acquisition time of 300 s.  Bias voltages $V$ refer to the sample voltage with respect to the tip.\\
\begin{figure}
\includegraphics[scale=0.5]{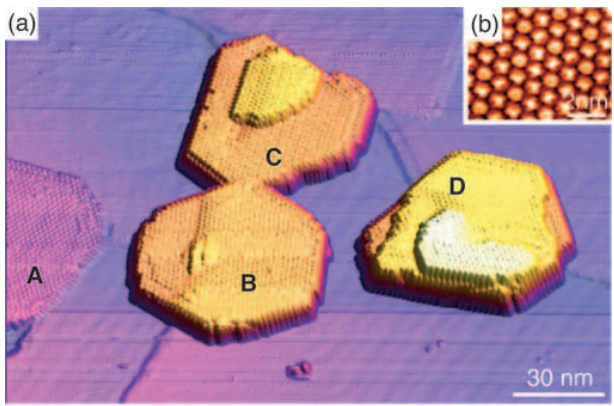}
\caption{\label{fig:figure1} (color online). (a) STM image of C$_{60}$ nanocrystals formed on a NaCl ultrathin film
grown on Au(111)  ($V\!=\!-3\,$V,  $I\!=\! 0.02\,$nA). Island \textbf{A} is a
C$_{60}$ monolayer on Au(111), the small blue triangle below  \textbf{A} is part of the bare Au surface. 
 Hexagonal island \textbf{B},
 and truncated
triangular islands \textbf{C}
and \textbf{D}
 consist of up to two, three, and
four C$_{60}$ molecular layers, respectively, on NaCl. (b)  Sub-molecular resolution on island \textbf{B} ($V\!=\!-3\,$V, $I\!=\! 0.1\,$nA).
}
\end{figure}
\indent NaCl  forms (100)-terminated islands on Au(111) of thickness between 1 and 3 monolayers  and width up to $1\ \mu$m. Contrary to the layer-by-layer growth of C$_{60}$ on Au(111) leading to extended islands found by STM \cite{Altman93}, on NaCl electron microscopy studies \cite{Saito92} revealed that the C$_{60}$ molecules aggregate into hexagonal or truncated triangular nanocrystals with a height of several molecular layers. This situation is well illustrated  in Fig.~\ref{fig:figure1}, where C$_{60}$ islands grown on both, the bare Au(111) (\textbf{A}) and the NaCl covered surface  (\textbf{B-D}) are visible.
The  nanocrystals present a minimum height of two layers of C$_{60}$ molecules (island \textbf{B}). The nucleation of the C$_{60}$  nanocrystals starts at defects of the NaCl layer (protrusions or vacancies),
 monatomic steps of Au(111) (covered with NaCl), or edges of the second layer of NaCl. As shown in Fig.~\ref{fig:figure1}(b), the C$_{60}$ molecules form hexagonally arranged layers with  an intermolecular distance of 1 nm.  \\
\begin{figure}
\includegraphics[scale=0.5]{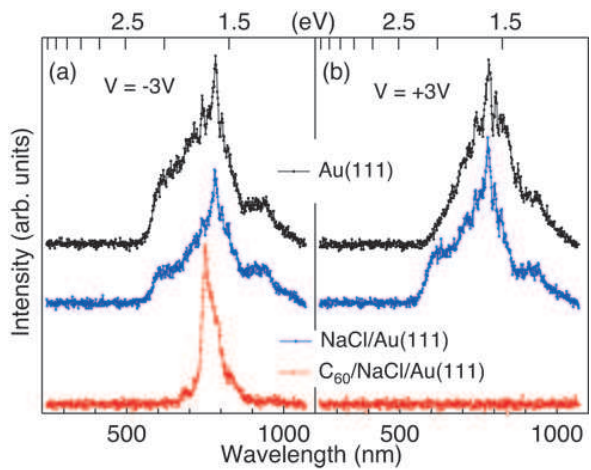}
\caption{\label{fig:figure2} (color online).  STM-induced photon emission spectra acquired  with  (a) negative ($V\!=\!-3\ $V, $I\!=\! 1\,$nA) and  (b) positive  ($V\!=\!+3\ $V, $I\!=\! 1\,$nA) bias polarity. Spectra acquired on bare Au(111) and on NaCl thin film reveal characteristic emission from a localized surface plasmon. Emission from C$_{60}$ can only be excited  at negative voltages.  Spectra are vertically shifted for clarity.
%
}
\end{figure}
\indent Figure~\ref{fig:figure2} shows STM-induced optical spectra from the bare Au(111) surface, the NaCl covered Au(111) surface and from a C$_{60}$ nanocrystal. Photons emitted from Au(111) originate from an IET process, involving excitation and decay of a surface  plasmon localized between the tip and the surface \cite{Berndt91,Berndt93B,Meguro02}.
A similar spectral shape is observed  over the NaCl layer, however with reduced  intensity due to the dielectric NaCl spacer layer.
Characteristic  for this process is the energy-dependent quantum cut-off \cite{Berndt91}   (not shown)
and the possibility to excite the emission with both bias polarities \cite{Berndt93B}, as shown in Fig.~\ref{fig:figure2}.
The  third spectrum in Fig.~\ref{fig:figure2}(a)
was acquired over a single C$_{60}$ molecule in the surface of a nanocrystal  ($V\!=\!-3\,$V,  $I\!=\! 1\,$nA).
Light emission from C$_{60}$ is observed only for  bias higher than the
threshold voltage of $V$ = $-$2.3 V.
The  emission onset is located at  $\approx\,$680\,nm  and its position is  independent of  the voltage.
For positive voltages up to $+$4.5 V  no photon emission is detected.
These observations clearly distinguish the light emission spectrum of C$_{60}$ from those acquired over the substrate (Au and NaCl).\\
\begin{figure}
\includegraphics[scale=0.5]{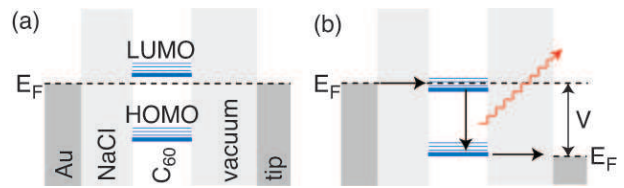}
\caption{\label{fig:figure3} (color online). Energy diagram of the double barrier tunneling junction at  (a) at zero-bias voltage,  (b) applied negative voltage, corresponding to the conditions for  luminescence. }
\end{figure}
\indent The occurrence of luminescence from the C$_{60}$ molecule is related to the  characteristics of the tunneling junction, as shown in Fig.~\ref{fig:figure3}.
At negative bias voltage larger than $-$2.3 V,
the highest-occupied molecular orbital (HOMO), which is completely filled for C$_{60}$ in the ground state,  is higher than the Fermi level ($E_F$) of the tip.
 The electrons are extracted from the HOMO and tunnel to the tip, while the lowest-unoccupied molecular orbital (LUMO), now lower than the Fermi level of the sample, is populated by the electrons  tunneling from the substrate, electrons that
 can radiatively decay into the partially empty HOMO (hot electron/hole injection).
The fact that luminescence is not observed for tunneling from the tip to the sample for voltages  up to  $+$4.5 V may be due to the asymmetry of the HOMO-LUMO gap with respect to  $E_F$ \cite{Lof92} and to the different properties of  the two tunneling barriers (vacuum and NaCl) \cite{Wu04}. \\
\begin{figure*}
\includegraphics[scale=0.5]{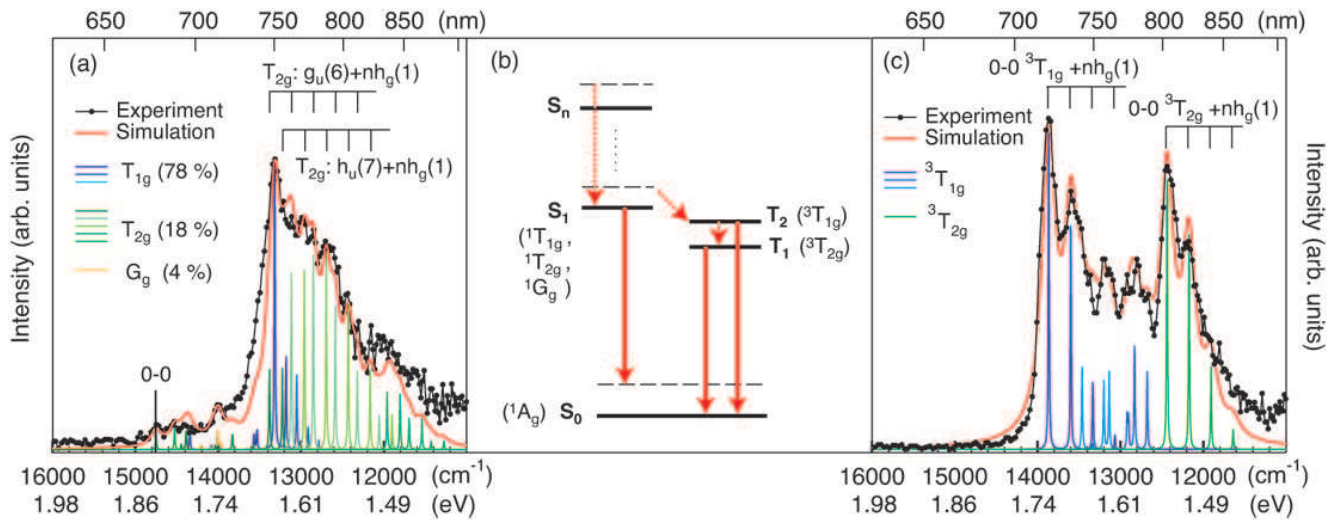}
\caption{\label{fig:figure4} (color online)  (a) STM-induced light emission spectrum assigned to
C$_{60}$  fluorescence  ($V\!=\!-3\,$V, $I\!=\! 1\,$nA) and calculated spectrum.
(b) Schematic diagram of the lowest singlet ($S_i$) and triplet ($T_i$) states. Horizontal solid lines: pure electronic levels; horizontal dashed lines:  vibrational  levels.
Solid  arrows represent electronic transitions;
dashed arrows represent radiationless mechanisms of relaxation
(internal conversion, intersystem crossing, vibrational relaxation).
(c) STM-induced light emission spectrum assigned to
C$_{60}$  phosphorescence  ($V\!=\!-3\,$V, $I\!=\!1\,$nA) and
calculated spectrum.
For both simulations, experimentally determined frequencies  for the vibronically induced  $S_0 \leftarrow S_1$ ($ T_{1g}$, $T_{2g}$, $G_{g}$) and $S_0 \leftarrow T_1$ ($^3T_{2g}$), $T_2$ ($^3T_{1g}$) transitions are used, see  Tab.~\ref{tab:table}  \cite{Sassara97,Schettino94}.
Each component has a Lorentzian lineshape, broadened by
150  cm$^{-1}$  (a) and  200 cm$^{-1}$ (c)  to obtain the calculated spectra (red).
}
\end{figure*}
\begin{table*}
\caption{\label{tab:table} HT and JT active modes used in the simulation of the STM-induced light emission spectra shown in  Fig.~\ref{fig:figure4}(a,c).  For the most intense modes contributing to the spectra, the experimental frequencies \cite{Sassara97,Schettino94} are indicated  in brackets.  }
\begin{ruledtabular}
\begin{tabular}{p{15mm}p{68mm}p{1.7mm}|p{1mm}p{16mm}p{68mm}}
\multicolumn{2}{l}{\textbf{Fluorescence}: $S_0 \leftarrow S_1$} &&&\multicolumn{2}{l}{\textbf{Phosphorescence}: $S_0 \leftarrow T_1, T_2$}\\
\hline\\
$S_1$ ($T_{1g}$)  & HT:  $t_{1u}(4)\,$(1430\,cm$^{-1}$),   $h_{u}(7)\,($1566\,cm$^{-1}$), $h_{u}(1)$,  $h_{u}(3)$,  $h_{u}(4)$,  $t_{1u}(3)$,  $h_{u}(5)$. JT:  $h_{g}(7)$,  $a_{g}(2)$, $h_{g}(1)$.  && & $T_1$ ($^3T_{2g}$) &  JT: $h_{g}(1)\,$(266\,cm$^{-1}$).\\
$S_1$ ($T_{2g}$) & HT: $g_{u}(6)\,$(1410\,cm$^{-1}$), $h_{u}(7)\,($1566\,cm$^{-1}$), $g_{u}(1)$,  $h_{u}(1)$,  $g_{u}(4)$,  $h_{u}(5)$. JT: $h_{g}(1)\,$(266\,cm$^{-1}$),  $h_{g}(7)$,  $a_{g}(2)$. &&&
 $T_2$ ($^3T_{1g}$)  &
HT: $t_{2u}(3)\,$(1037\,cm$^{-1}$), $h_{u}(1)$,   $h_{u}(2)$,   $h_{u}(4)$,  $a_{u}(1)$,  $g_{u}(4)$,  $t_{1u}(3)$. JT: $h_{g}(1)\,$(266\,cm$^{-1}$). \\
$S_1$ ($G_{g}$)  &  HT: $h_{u}(4)\,$(738\,cm$^{-1}$), $h_{u}(2)$, $h_{u}(3)$,  $g_{u}(3)$, $g_{u}(5)$, $t_{2u}(2)$.&&&\\
\end{tabular}
\end{ruledtabular}
\end{table*}
%
%
\indent  Figure~\ref{fig:figure4}(a) shows the same luminescence spectrum  obtained  from  an individual  C$_{60}$ molecule as in Fig.~\ref{fig:figure2}(a), but corrected for the quantum efficiency of the detection system.
 In order to identify the electronic and vibronic transitions giving rise to the observed emission, we  compare our results with laser-induced high-resolution photoluminescence data \cite{Sassara96B,Sassara97} and with quantum chemical calculations \cite{Sassara97,Orlandi02}. It is now established  that the lowest excited singlet state $S_1$  has mixed $T_{1g}$, $T_{2g}$ and $G_{g}$ character \cite{Sassara97}. The electric dipole transitions from this state to the ground state $S_0$ ($A_{g}$) are symmetry forbidden, but they occur through  HT and JT electron-vibration coupling mechanisms of intensity borrowing \cite{Negri92,Orlandi02}.
The relaxation of the selection rules due to symmetry lowering in the C$_{60}$ lattice
gives rise to a very weak luminescence signal corresponding to the pure electronic (0-0)  $S_0 \leftarrow S_1$ transition, found at $\approx\,$678 nm, as indicated in
Fig.~\ref{fig:figure4}(a).
The  red shift of about 40 nm with respect to C$_{60}$ in the  gas phase is  attributed to environmental effects \cite{Sassara96B,Sassara97,vandenHeuvel95}.
The  observation of the pure electronic origin helps to determine  the  vibronically induced false origins  as in the high-resolution photoluminescence measurements
\cite{Hung96,Sassara97}.
The fluorescence spectra are simulated using calculated oscillator strengths and experimentally determined frequencies for the HT vibronically induced  $S_0 \leftarrow S_1$ ($ T_{1g}$, $T_{2g}$, $G_{g}$) transitions, and the experimental frequencies for the JT  active modes, as presented in Fig.~\ref{fig:figure4}(a) and in  Tab.~\ref{tab:table}  \cite{Sassara97,Schettino94}.
The contribution of each  symmetry character of $S_1$ varies slightly from one probed molecule to another, reflecting the known sensitivity of  C$_{60}$ to the local environment \cite{Sassara97,vandenHeuvel95B,Chergui05}.
The agreement between measured and calculated spectra in Fig.~\ref{fig:figure4}(a)  demonstrates the local character of the measurement and provides evidence for the preservation of  the  C$_{60}$  molecular properties  in the van de Waals crystal, characterized by weak interactions between the molecules. \\
\indent Interestingly, we also observe another type of  electronic transitions,  shown in Fig.~\ref{fig:figure4}(c). Similar  spectra have been reported
for  laser-induced luminescence from solid  C$_{60}$ \cite{Guss94,vandenHeuvel95B,Akimoto02}, and have recently been identified as phosphorescence originating from   triplet  to  singlet ground state transitions  \cite{Chergui05}.
Although symmetry and spin-forbidden,
intense pure electronic  (0-0) triplet to singlet transitions  have been observed  in C$_{60}$ phosphorescence spectra \cite{Sassara96,vandenHeuvel94}.
The low-energy part of the spectrum in Fig.~\ref{fig:figure4}(c) arises from  the  $S_0 \leftarrow  T_1$ ($^3T_{2g}$) transition, characterized by the intense 0-0 origin  at
$\approx\,$803 nm and by the progression of a JT active mode,  see Tab.~\ref{tab:table},
 in agreement with phosphorescence spectra obtained from dispersed   C$_{60}$  molecules  \cite{vandenHeuvel94,Sassara96,Hung96}.
The electronic origin is shifted by 27 nm to the red with respect to the estimated gas phase energy   \cite{Sassara96,Chergui05}.
The high-energy region of the spectrum presents a similar shape, i.e. an intense transition, located at  $\approx\,$720 nm, and a progression of vibronic bands.
The observed energy difference   of $\approx\,$0.18 eV between the two most intense features in the spectrum shown in Fig.~\ref{fig:figure4}(c)   is in agreement with electron energy loss spectroscopy results \cite{Gensterblum91,Chergui05} and with calculations \cite{Laszlo89} for the splitting of the lowest C$_{60}$ triplet states.
Therefore, the high-energy part of the spectrum in Fig.~\ref{fig:figure4}(c) is
assigned  to  transitions from the next higher triplet state $S_0 \leftarrow T_2$ ($^3T_{1g}$) \cite{Chergui05}.
The good agreement between the calculated and the measured spectra allows us  to assign the STM-induced phosphorescence spectrum to light emission from an individual
C$_{60}$ molecule, as in the case of fluorescence.
This finding  contradicts previous interpretations of similar spectra obtained by laser induced luminescence from solid C$_{60}$ in terms of  excitonic emission  \cite{Guss94} or emission delocalized over more than one  C$_{60}$ molecule \cite{vandenHeuvel95B,Akimoto02}.\\
\indent The observation of both radiative relaxation processes, fluorescence and phosphorescence, in the STM-induced light emission may be related to (i) the sensitivity of the probed  C$_{60}$ molecule to the local environment  in the nanocrystal and/or to (ii) the actual tunneling conditions.
(i) Differences in the local C$_{60}$ environment may induce a modification of the mixed character of the states and/or a relaxation of the selection rules,
as observed in ensemble-averaged experiments  \cite{Hung96,Sassara96,Sassara96B,Sassara97,vandenHeuvel94,vandenHeuvel95,
Guss94,vandenHeuvel95B,Akimoto02,Chergui05}.
 Novel, here, is the fact that the influence of the environment is probed on an individual selected molecule.
(ii) Even for equal nominal tunneling parameters, the actual tunneling conditions can vary from one measurement to the other.  Tip shape and composition, current instabilities, or electric field fluctuation  may influence the relaxation paths, for example by enhancing the intersystem crossing, and increasing the population of the triplet states.
The identification of the physical origin of the observation of both, fluorescence and phosphorescence calls for future time-resolved luminescence studies employing both, non-local laser excitation and local STM-induced excitation of supported molecules.\\
\indent To summarize, unambiguous chemical identification of individual C$_{60}$ molecules is obtained via their luminescence
 induced by tunneling electrons.
 Emission from three electronic states mapping more than twenty vibrational
 Jahn-Teller and Herzberg-Teller
 modes of the molecule is identified,  in excellent agreement with the known energies of the electronic \cite{Hung96,Sassara96,Sassara96B,Sassara97,vandenHeuvel94,Chergui05} and vibrational \cite{Sassara97,Orlandi02,Schettino94} levels  of  C$_{60}$.
The present observation of local  fluorescence and phosphorescence demonstrates
the capability of STM-induced light emission  for the chemical recognition on  the single-molecular scale.\\
\indent The financial support of the Swiss National Science Foundation is acknowledged.

\end{document}